\documentclass[aps,prl,twocolumn,superscriptaddress,showpacs,]{revtex4-1}

\usepackage{amsmath}
\usepackage{amssymb}
\usepackage{graphics}
\usepackage{graphicx}
\usepackage{epstopdf}
\usepackage{dcolumn}
\usepackage{bm}
\usepackage{xcolor}
\usepackage[marginal]{footmisc}
\usepackage[marginal]{footmisc}
\usepackage{flushend}
\usepackage{hyperref}
\usepackage{natbib}
\hypersetup{hidelinks}
\begin{document}

\title{Regular and irregular revivals of quasi-periodic random waves}
\author{Yanlin Bai}
\affiliation{Shandong Provincial Engineering and Technical Center of Light Manipulation \& Shandong Provincial Key Laboratory of Optics and Photonics Devices, School of Physics and Electronics, Shandong Normal University, Jinan 250014, China}
\affiliation{Collaborative Innovation Center of Light Manipulations and Applications, Shandong Normal University, Jinan 250358, China}
\author{Yangjian Cai}
\email[]{yangjian\_cai@163.com}
\affiliation{Shandong Provincial Engineering and Technical Center of Light Manipulation \& Shandong Provincial Key Laboratory of Optics and Photonics Devices, School of Physics and Electronics, Shandong Normal University, Jinan 250014, China}
\affiliation{Collaborative Innovation Center of Light Manipulations and Applications, Shandong Normal University, Jinan 250358, China}
\author{Chunhao Liang}
\email[]{chunhaoliang@sdnu.edu.cn}
\affiliation{Shandong Provincial Engineering and Technical Center of Light Manipulation \& Shandong Provincial Key Laboratory of Optics and Photonics Devices, School of Physics and Electronics, Shandong Normal University, Jinan 250014, China}
\affiliation{Collaborative Innovation Center of Light Manipulations and Applications, Shandong Normal University, Jinan 250358, China}
\author{Sergey A. Ponomarenko}
\email[]{serpo@dal.ca}
\affiliation{Department of Electrical and Computer Engineering, Dalhousie University, Halifax, Nova Scotia, B3J 2X4, Canada}
\affiliation{Department of Physics and Atmospheric Science, Dalhousie University, Halifax, Nova Scotia, B3H 4R2, Canada}

\date{\today}

\begin{abstract}
Paraxial wave packets with discrete spatial, temporal, or spatiotemporal spectra are known to undergo periodic axial revivals on propagation in either free space or linear transparent, weakly dispersive media. Such spectacular revivals, ubiquitously encountered in physics, from optics and acoustics to condensed matter physics, are distinguished by their strict periodicity. We show theoretically and verify experimentally that ensembles of quasi-periodic random wave packets exhibit a unique revival network composed of regular (periodic) and irregular (aperiodic) revivals. Moreover, individual realizations of a statistical ensemble self-reconstruct, in general, at different propagation distances than do ensemble averages. Our results shed new light on the fundamental physics of self-reconstruction of random wave packets with structured correlations. 
\end{abstract}
\maketitle
\textit{Introduction---}The classic Talbot effect manifests periodic axial revivals of a coherent paraxial wave packet diffracted by a grating, as the packet self-reconstructs past the grating over multiples of the  Talbot distance~\cite{talbot1836facts}.  The effect has played a fundamental role in wave physics ever since its theoretical explanation by Lord Rayleigh~\cite{rayleigh1881xxv} due, in large part, to its ubiquitous nature: Talbot revivals have been discovered for electromagnetic waves spanning a wide range of frequencies from optical to x-rays~\cite{patorski1989self}, as well as for matter~\cite{deng1999temporal} and even for water~\cite{bakman2019observation} waves. In addition, veiled spatial revivals~\cite{yessenov2020veiled} and the Talbot effect in space and time~\cite{hall2021space} have also been demonstrated for spatiotemporal wave packets. At the same time, Talbot or generic wave revivals have found numerous applications, including in lensless imaging~\cite{wen2013talbot}, fluorescence microscopy \cite{han2013wide}, temporal cloaking~\cite{li2017extended}, phase locking of laser arrays~\cite{tradonsky2016talbot}, and prime number decomposition~\cite{bigourd2008factorization,pelka2018prime,liu2023axial}. 

To date, the vast majority of research on the Talbot effect has dealt with revivals of deterministic wave packets that are fully spatially and temporarily coherent~\cite{wen2013talbot}. Although revivals of incoherent waves were also examined by Lau~\cite{lau1948beugungserscheinungen}, there has been surprisingly little work on self-reconstruction of random wave packets which are, in general, \emph{partially coherent}. Such partially coherent wave packets possess a unique degree of freedom, the \emph{degree of coherence} (DOC), which quantifies the extent of second-order correlations of the wave packet fields at pairs of points. Structuring the DOC at the source has enabled a number of seminal advances in imaging through random environments~\cite{clark2012high,pascucci2019compressive,katz2014non,xu2022structurally}, optical communications and cryptography~\cite{peng2021optical,liuunlocking,lu2024coherence}, metrology~\cite{batarseh2018passive}, and optical computing~\cite{liu2023axial,dong2024partial}. Moreover, a recent work on diffraction of partially coherent beams by a grating has revealed the emergence of intricate coherence needles in the DOC of the diffracted beam field~\cite{schouten2025gratings}. However, a diffraction grating imposes the same periodicity on the diffracted partially coherent field and its second-order correlations, and hence the corresponding DOC. Thus, to the best of our knowledge, the revivals of random wave packets with \emph{independently} periodically structured DOC have not been explored to date.  This raises a fundamental question: How does the commensurability or incommenusurability of the periods of the field intensity and degree of coherence of a random wave packet affect the wave packet revivals?

In this work, we elucidate this point by studying the revivals of random wave packets whose periodic intensities and DOCs possess different periods. We demonstrate that, counterintuitively, even if the periods of intensity and DOC are not equal but commensurable, implying strict periodicity of any ensemble realization, it takes, in general, longer distances for the individual realizations of the ensemble to self-reconstruct than it does for the corresponding ensemble averages. Moreover, ensemble averaged revivals manifest features that, to our knowledge, have no parallels in coherent wave packet revivals. We also show that the incommensurability of the intensity period and the degree-of-coherence period results in a unique revival network, or Talbot carpet, composed of regular (periodic) and irregular (aperiodic) revivals.

We support our theoretical insights with experiments.  Although we explore theoretically the revivals of ultrashort pulses in dispersive media, our findings are equally valid for paraxial beams or quantum wave packets self-reconstructing in free space. To bring this point home, we performed our experiments with paraxial beams. Thus, our results point to a fundamental link between the revival network architecture of generic random wave packets with periodic second-order correlations and rational/irrational numbers, which are objects of much interest in number theory. 

\textit{Theoretical framework---}The evolution of an ensemble of random envelopes $\{\Psi\}$ of optical pulses in a linear dispersive medium is governed by the Schr\"{o}dinger-like equation
\begin{equation}\label{LSE}
    i\partial_z\Psi-\textstyle\frac{\beta_2}{2}\partial_{\tau\tau}^2 \Psi=0.
\end{equation}
Here $\tau=t-\beta_1 z$ is a retarded time in the reference frame moving with the group velocity $v_g=\beta_1^{-1}$ of the pulse and $\beta_2$ is a group velocity dispersion coefficient. We note that owing to the space-time duality \cite{kolner2002space,salem2013application}, paraxial sheets of light ($1+1$D beams) propagating in free space obey Eq.~(\ref{LSE}) as well. 

We consider a random pulse source of the form
\begin{equation}\label{Psi}
    \Psi(t,0)=\alpha(t)\psi_0 (t),
\end{equation}
where $\alpha$ is a complex random amplitude (multiplicative noise), which may be, for example, a random phasor, $e^{i\Phi}$, and $\psi_0$ is a deterministic periodic pulse envelope with a period $T_p$. We assume $\alpha(t)$ to be statistically stationary and periodic with a period $T_c$, such that we can expand it into a Fourier series as
\begin{equation}\label{a}
    \alpha(t)=\sum_{m=-\infty}^{\infty} a_m e^{-i2\pi m t/T_c}.
\end{equation}
Here the (random) amplitudes $\{a_m\}$ of the modes are uncorrelated, so that
\begin{equation}\label{stat}
    \langle a_m^* a_n \rangle=\lambda_m \delta_{mn},
\end{equation}
where the angle brackets denote ensemble averaging. It follows at once from Eqs.~(\ref{Psi}) through~(\ref{stat}) and Eq.~(\ref{G}) that the DOC at the source, defined as $\gamma_0 (t_1,t_2)=\Gamma(t_1,t_2,0)/\sqrt{\Gamma (t_1,t_1,0)\Gamma (t_2,t_2,0)}$~\cite{mandel1995optical}, becomes
\begin{equation}\label{OCL}
    \gamma_0(t_1-t_2)=\sum_{m=-\infty}^{\infty} \nu_m e^{-i2\pi m(t_2-t_1)/T_c}.
\end{equation}
Here $\nu_m=\lambda_m/\sum_m \lambda_m$ is a normalized power spectrum of field correlations at the source. Eq.~(\ref{OCL}) describes a temporal optical coherence lattice (OCL) introduced in~\cite{ma2014optical}; spatial OCLs were theoretically discussed~\cite{ma2015free} and generated in the laboratory~\cite{liang2021perfect}. 

It follows from Eqs.~(\ref{LSE}) through (\ref{a}) (Sec. S1, Supplemental Material~\cite{supplemental}) that any statistical realization of the ensemble $\{\Psi\}$ can be represented at any $z\geq 0$ as
\begin{equation}\label{APS}
\begin{split}
     \Psi(\tau,z)&=\sum_{m,n=-\infty}^{\infty}a_m c_n e^{-2\pi i(n/T_p+m/T_c)\tau} \\
     &\times e^{-i2\pi^2\beta_2 z(n/T_p+m/T_c)^2}.
\end{split}
\end{equation}
where $\{c_n\}$ are Fourier series coefficients of $\psi_0$.  At the source, $z=0$, Eq.~(\ref{APS}) describes a quasi-periodic function of time~\cite{amidror2009theory}, which is a particular case of a Bohr almost periodic function~\cite{bohr2018almost}. The structure of revivals governed by Eq.~(\ref{APS}) is crucially dependent on the relationship between $T_p$ and $T_c$. 

\textit{Commensurable $T_p$ and $T_c$---}We first explore the case
\begin{equation}\label{TpTc}
    T_c/T_p=p/q,
\end{equation}
where $p$ and $q$ are coprime. It follows from Eq.~
(\ref{TpTc}) that there exists a common period $T=pT_p=qT_c$, such that any ensemble realization is strictly periodic. Since each realization corresponds to a specific set $\{a_m\}$, it manifests classic Talbot revivals (Fig.S1, Supplemental Material~\cite{supplemental}) with the self-imaging distance~\cite{patorski1989self} 
\begin{equation}\label{Lin}
    L_{\mathrm{SI}}=\frac{T^2}{\pi|\beta_2|}.
\end{equation}

Next, we show in the Supplemental Material~\cite{supplemental} that the autocorrelation function of the ensemble $\{\Psi\}$, given explicitly by Eq.~(\ref{Gam}), self-reconstructs over multiples of $L_{\mathrm{SI}}$, while replicas of $\Gamma(t_1,t_2,0)$, delayed or advanced by $T/2$ depending on the sign of $\beta_2$, emerge over multiples of $L_{\mathrm{SI}}/2$. However, more instructively, the analysis of Eq.~(\ref{Gam}) uncovers the existence of another set of perfect revivals at  multiples of the distance
\begin{equation}\label{Len}
    \overline{L}_{\mathrm{SI}}=
\begin{cases}
    \frac{qT_pT_c}{2\pi|\beta_2|}, & p=2f; \\
    \frac{qT_pT_c}{\pi|\beta_2|}, & p=2f-1,
\end{cases}
\end{equation}
where $f\in\mathbb{N}$. We refer to $\overline{L}_{\mathrm{SI}}$ as an \emph{ensemble self-imaging} distance. A quick look at Eqs.~(\ref{TpTc}) through~(\ref{Len}) reveals that $\overline{L}_{\mathrm{SI}}\leq L_{\mathrm{SI}}$ with the equality taking place only if $p=1$, implying that $T=T_p$. This result seems counterintuitive at first glance. To explain unexpected early revivals of ensemble averages, we first notice that, in general, a large number of Fourier harmonics, comprising the source field, are required to get in phase for complete self-reconstruction of any ensemble realization at a given propagation distance. However, according to Eq.~(\ref{stat}), all $\{a_m\}$ are uncorrelated. Therefore, if second-order source correlations exhibit periodicity, only the modes $\{c_n\}$, which make up the deterministic pulse envelope, should synchronize to ensure the revival of the ensemble autocorrelation function. The latter, in general, occurs over shorter distances than the phase locking of all harmonics required to reconstruct any individual realization of the ensemble. 

Our analysis (Sec. S2, Supplemental Material~\cite{supplemental}) indicates that, in general, revivals do not depend on the power spectrum $\nu_m$ of random modes. However, to illustrate our results, we hereafter choose a (normalized) Gaussian power spectrum $\nu_m=e^{-2\xi_c m^2}$ with $\xi_c=0.1$, where $\xi_c$ is a coherence parameter. We note that $\xi_c\rightarrow\infty$ corresponds to a fully coherent wave packet with only the lowest-order mode $m=0$ present in Eq.~(\ref{OCL}), implying that $|\gamma_0 (t_1,t_2)|=1$. 

\begin{figure}[t]
\centering
\includegraphics[width =\linewidth]{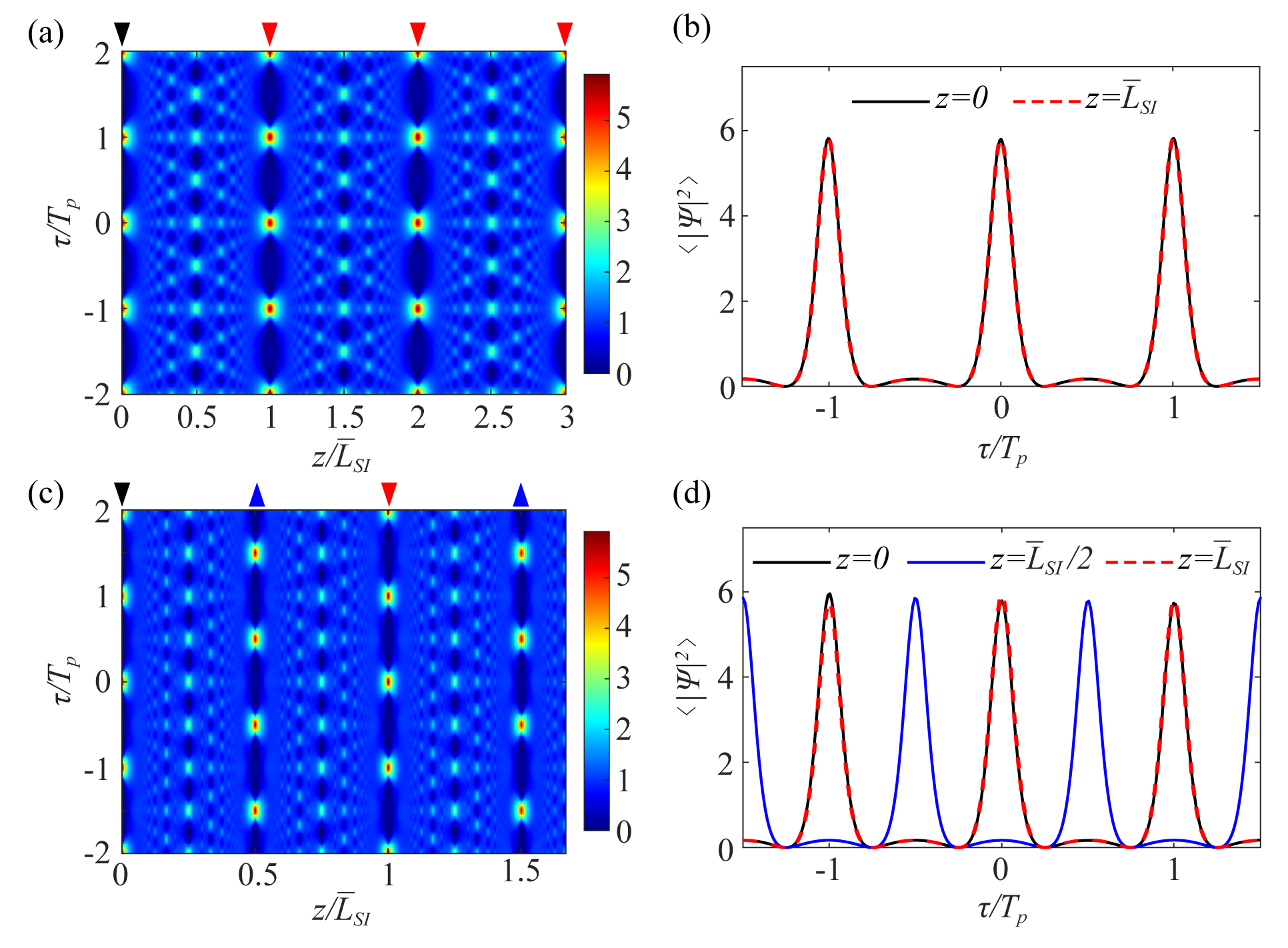}
\caption{(a) \& (c) Talbot carpet of an ensemble averaged intensity corresponding to (a) $q=1$ and $p=2$, as well as  (c) $q=1$ and $p=3$. The inverted black and red triangles mark the intensity at the source and its perfect replicas at multiple integers of $\overline{L}_{\mathrm{SI}}$. The upright blue triangles mark the replicas of the source shifted by half the period $T_p$. (b) \& (d) Intensity profile at the source (solid black) and replicas of the source intensity at $z=\overline{L}_{\mathrm{SI}}/2$ (solid blue) and at $z=\overline{L}_{\mathrm{SI}}$ (red dash). We employ a (normalized) Gaussian mode power spectrum $\nu_m=e^{-2\xi_c m^2}$ with $\xi_c=0.1$. } 
\label{fig1}
\end{figure}

We show a network of revivals of the average intensity of the source field in Fig.~\ref{fig1}(a)\&(c). Henceforth, we assume anomalous dispersion. We mark averaged source intensities and their replicas at $\overline{L}_{SI}$ with inverted black and red triangles, respectively. At the same time, shifted replicas are indicated by upright blue triangles. We exhibit the corresponding intensity profiles in Figs.~\ref{fig1}(b) \& ~\ref{fig1}(d). We can infer from the figure that revivals are strongly affected by the parity of $p$. While the Talbot carpet for odd $p$ involves replicas of the source at multiples of $z=\overline{L}_{SI}$ and those shifted by half the period $T_p$ of the deterministic pulse amplitude at the source at odd multiples of $z=\overline{L}_{SI}/2$, the revival network for even $p$ does not contain any shifted replicas of the source (see Figs.\ref{fig1}(b) and ~\ref{fig1}(d)). 

Our further analysis (Sec. S2, Supplemental Material~\cite{supplemental}) shows the absence of revivals at half the ensemble self-imaging distance for even $p$ to be a generic feature of the Talbot carpet of periodic random wave packets. We illustrate this general conclusion with a Talbot carpet for the average intensity corresponding to $p=2$ in Fig. S2. However, if the power spectrum $\lambda_m$ of the source field correlations contains no odd $m$ components, perfect replicas of $\Gamma(t_1,t_2,0)$, shifted by a quarter period $T$, emerge at $z=\overline{L}_{SI}/2$ (and odd multiples thereof) for even $p$ (see Sec. S2 and Fig S3 of the Supplemental Material~\cite{supplemental}).  We stress that the profound dependence of random wave revivals on the parity of $p$ and their sensitivity to the parity of power spectrum correlations of the source---which, to our knowledge, have not been reported to date---have no peers in any scenario of coherent wave packet revivals. 

\begin{figure}[t]
\centering
\includegraphics[width =\linewidth]{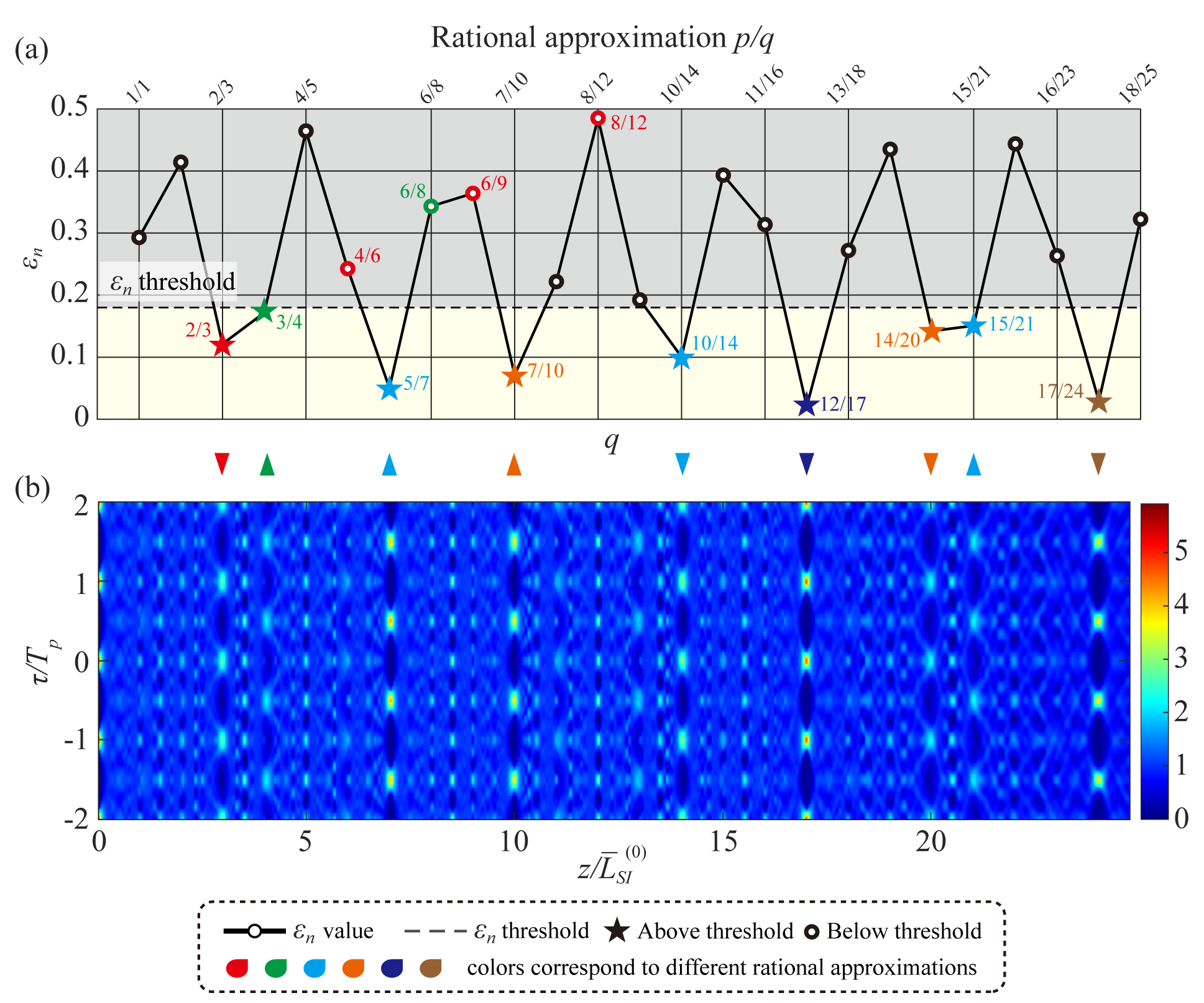}
\caption{(a) Rational approximation of $\alpha=T_c/T_p=1/\sqrt{2}$ as a function of $q$. (b) Talbot carpet of the average intensity for $\alpha=1/\sqrt{2}$ as a function of the scaled propagation distance $z/\overline{L}_{\mathrm{SI}}^{(0)}$. Acceptable candidates for complete revivals are marked by colored inverted triangles, while shifted acceptable revivals are denoted by colored upright triangles. We use a (normalized) Gaussian mode power spectrum $\nu_m=e^{-2\xi_c m^2}$ with $\xi_c=0.1$. } 
\label{fig2}
\end{figure}

\textit{Incommensurable $T_p$ and $T_c$---}In this case, the source field is quasi-periodic and, strictly speaking, all revivals are fractional. However, by approximating an irrational ratio $\alpha=T_c/T_p$ by a rational number $p/q$, such that the error $\varepsilon_n$,
\begin{gather}\label{eps}
    \varepsilon_n=|nq' T_c/T_p-np'|=n\varepsilon; \qquad n\in \mathbb{N},
\end{gather}
is small enough, we can make any replica of the source as close to perfect as we wish within a desired tolerance determined by the error. Here $p'$ and $q'$ are coprime, and $n=\mathrm{gcd}(p,q)$ is the greatest common divisor of $p$ and $q$. In Fig.~\ref{fig2}, we show the Talbot carpet for a representative case of $\alpha=1/\sqrt{2}$. We scale the longitudinal distance $z$ to $\overline{L}_{\mathrm{SI}}^{(0)}=T_cT_p/(2\pi|\beta_2|)$, which corresponds to the shortest available ensemble self-imaging distance in Eq.~(\ref{Len}).
In Fig.~\ref{fig2}a, we graph the error as a function of $q$. The associated self-reconstruction quality in Fig.~\ref{fig2}b is closely related to the error of a rational approximation: the smaller the error, the closer the intensity of a replica to the (average) source intensity. More precisely, we quantify the quality of self-imaging by a figure-of-merit that we refer to as fidelity (Sec. S4, Supplemental Material~\cite{supplemental}). This allows us to introduce a quantitative threshold that separates acceptable from poor replicas of the source intensity. In our representative example, $p/q=3/4$ is a marginal approximation which falls right at the threshold. Next, we mark acceptable and poor source replicas, corresponding to different rational approximations of $1/\sqrt{2}$, by pentagrams and open circles in Fig.~\ref{fig2}a, respectively. Furthermore, we indicate the corresponding acceptable revivals in Fig.~\ref{fig2}b with variable colors. Just as in Fig.~\ref{fig1}, the inverted triangles mark (nearly) perfect replicas of the source, while upright triangles denote shifted images. We can infer from Fig.~\ref{fig2} that all acceptable replicas are located at the following distances from the source:
\begin{equation}\label{Len-in}
    L_{q_{\varepsilon}}=\frac{q_{\varepsilon}T_c T_p}{2\pi|\beta_2|},
\end{equation}
where $q_{\varepsilon}\in\mathbb{N}$ corresponds to an acceptable image determined by the tolerance in Eq.~(\ref{eps}). 

The analysis of Fig.~\ref{fig2}b reveals a striking feature: the Talbot carpet is comprised of two types of revival networks. The first type contains uniformly spaced (regular) replicas of the source. Regular revivals can be associated with sets of reducible fractions $\{5/7,10/14,15/21\}$ (light blue triangles), or with mixed reducible and irreducible fractions, for instance, $\{10/14,12/17,14/20\}$. At the same time, all regular revival networks are embedded into a higher-level network, composed of all acceptable revivals (triangles of all colors) that are irregular as there is no clear pattern to the distance between adjacent revivals. In this context, it follows from Eq.~(\ref{eps}) that $\varepsilon_n$ increases with $n$. This observation explains the progressive loss of self-imaging quality of higher-order regular revivals in Fig.~\ref{fig2}b (see also Sec. S3 and Fig.S4 of the Supplemental Material~\cite{supplemental}). 

\textit{Experimental results---} Inspired by the space-time duality \cite{kolner2002space,salem2013application}, we designed and built an optical system capable of synthesizing spatially periodic or quasi-periodic random light fields. The latter are generated by illuminating a spatial light modulator, loaded with a set of predesigned holograms, with a laser light source with the carrier wavelength $\lambda = 632.8$ nm (see Sec.S4 of the Supplemental Material~\cite{supplemental} for details of our experimental setup and protocol). In our experiments, we choose a deterministic amplitude envelope of the source to be a simple cosine grating, such that  $\psi_0 (x)\propto 1+u\cos(2\pi x/\sigma_p)$, with a unit depth of modulation, $u=1$, and a period $\sigma_p=0.128$ mm. The DOC of our spatial OCL is given by Eq.~(\ref{OCL}) with a spatial period of $\sigma_c$ and the same normalized Gaussian power spectrum of mode correlations, $\nu_m=e^{-2\xi_c m^2}$, $\xi_c=0.1$.

\begin{figure}[t]
\centering
\includegraphics[width =\linewidth]{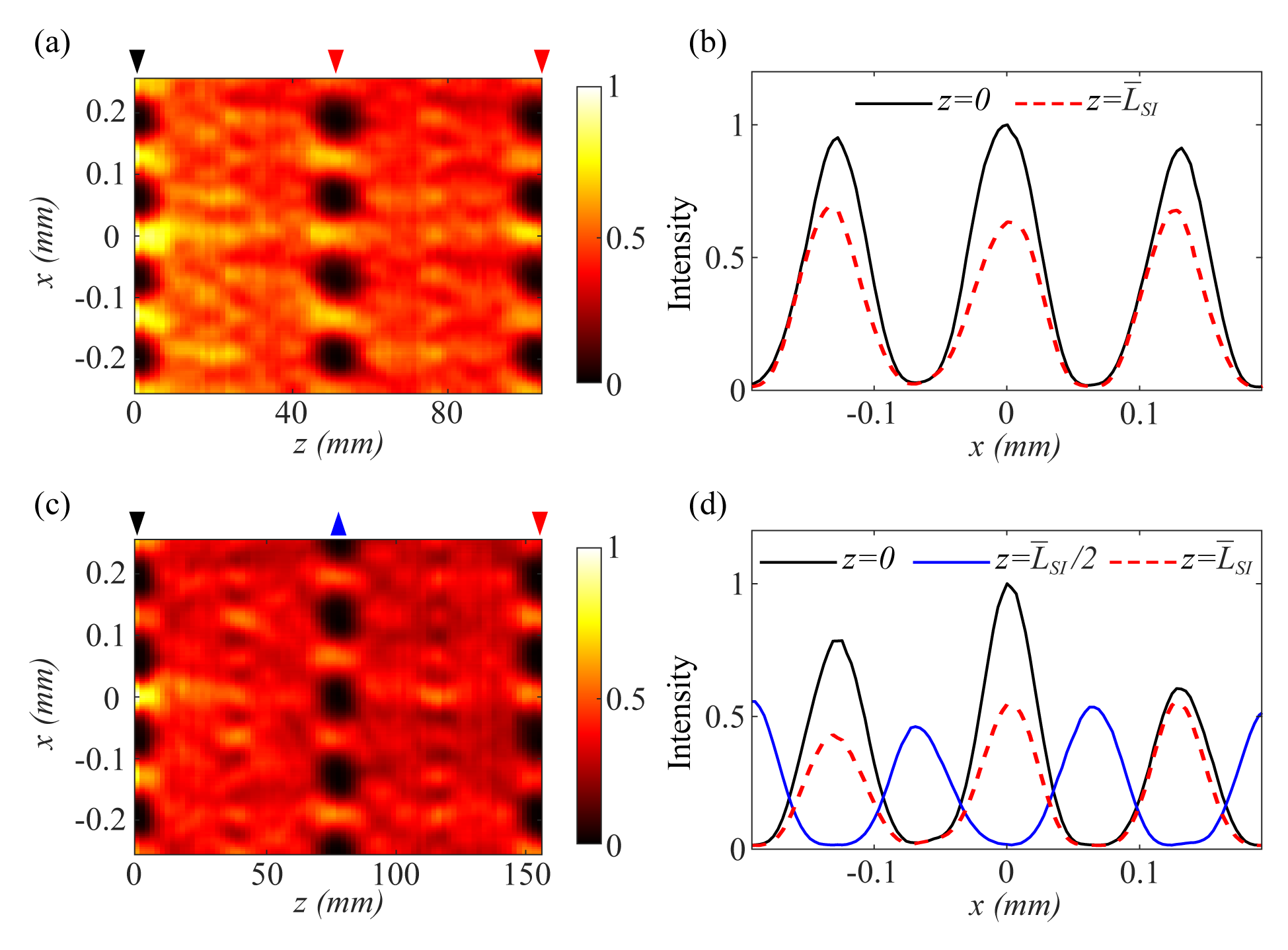}
\caption{Experimentally generated Talbot carpets for (a): $q=1$, $p=2$, and (c): $q=1$, $p=3$. The inverted black and red triangles mark the intensity at the source and its replica, at the self-imaging distance. The upright blue triangles indicates a half-period shifted replica at half the self-imaging distance. (b) \& (d) Intensity profile at the source (solid black) and replicas of the source intensity at half- (solid blue) and full (red dash) self-imaging distances. We employ a (normalized) Gaussian mode power spectrum $\nu_m=e^{-2\xi_c m^2}$ with $\xi_c=0.1$.} 
\label{fig3}
\end{figure}

In Fig. \ref{fig3}(a) and 3(c), we show  our experimental results for ensemble-averaged Talbot carpets corresponding to the source with commensurable $\sigma_p$ and $\sigma_c$, $\sigma_c/\sigma_p=p/q$ with $q=1$, for (a) $p=2$ and (c) $p=3$. The inverted red triangles mark perfect images of the average intensity of the source, while the upright blue triangles indicate replicas shifted by half a period. We show the associated intensity profiles at selected propagation distances in Figs.\ref{fig3}(b) and 3(d). We can infer from the figure that our measurements qualitatively confirm a strong dependence of the structure of the revival network on the parity of $p$, in complete agreement with the theoretical predictions. However, the peak amplitudes of the replicas are attenuated compared to those of the source. To elucidate this point, we notice that replicas of any individual ensemble realization show excellent quantitative agreement with the source (Fig S8, Supplemental Material~\cite{supplemental}). However, even tiny errors during the hologram refresh, or those induced by mechanical vibrations of our setup, can amplify significantly upon averaging over many ensemble realizations, thus yielding quantitative discrepancies. 

\begin{figure}[t]
\centering
\includegraphics[width =\linewidth]{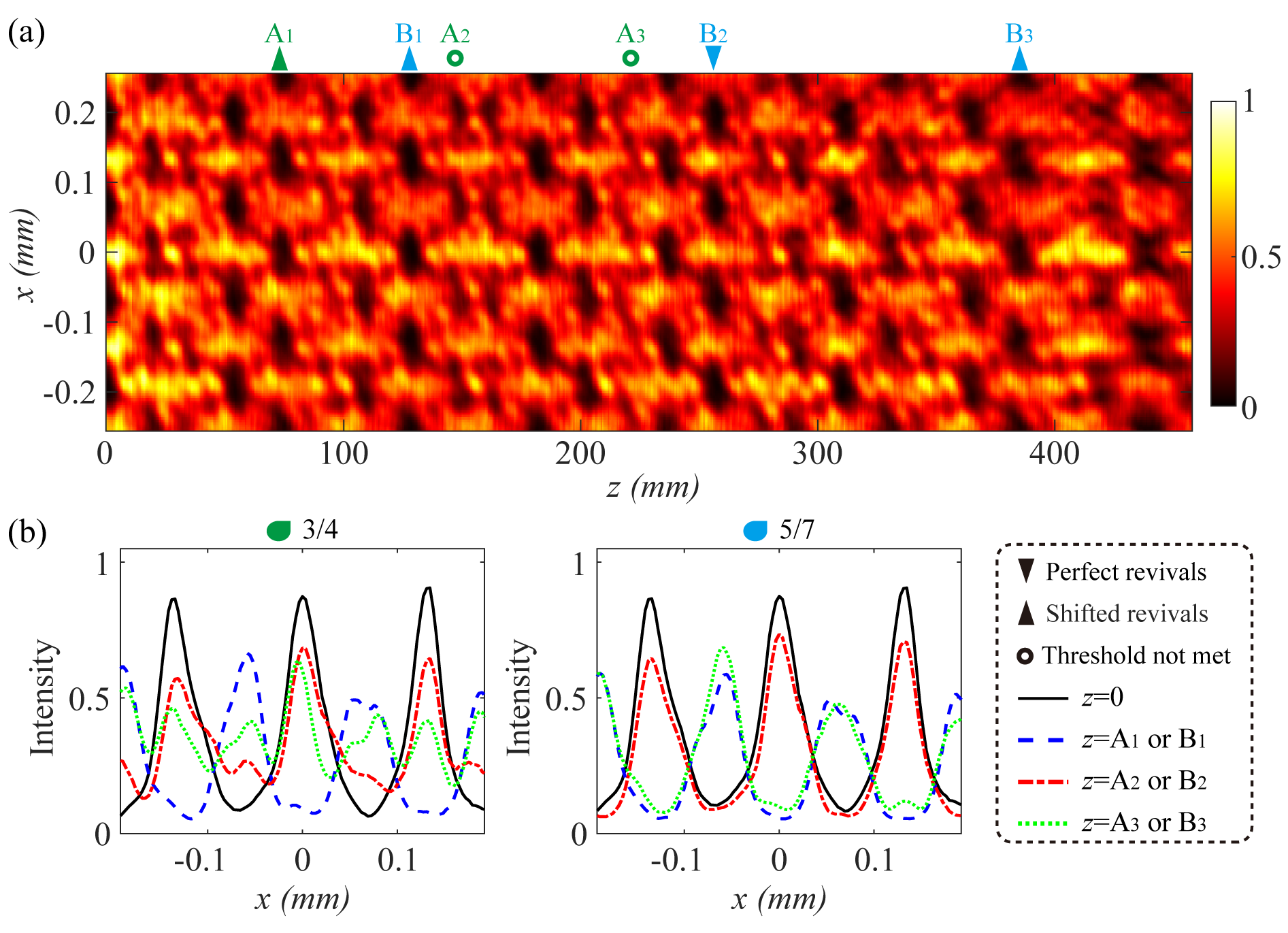}
\caption{(a) Experimentally realized Talbot subnetwork that includes revivals associated with rational approximations $3/4$ (green) and $5/7$ (blue) of $1/\sqrt2$. Upright green and blue triangles mark shifted images, while the inverted blue triangle denotes an image of the source. (b) Intensity profiles of the source and those of images of variable order corresponding to the same rational approximations as in panel (a). We use a (normalized) Gaussian mode power spectrum $\nu_m=e^{-2\xi_c m^2}$ with $\xi_c=0.1$.} 
\label{fig4}
\end{figure}

We illustrate our results for the quasi-periodic case, $\sigma_c/\sigma_p=1/\sqrt2$ in Fig.\ref{fig4}. We note that in both Fig.~\ref{fig3} and Fig.~\ref{fig4}, the ensemble self-imaging distance $\overline{L}_{\mathrm{SI}}$ is given by Eqs.~(\ref{Len}) and~(\ref{Len-in}), respectively, with the temporal periods replaced with the spatial ones and $|\beta_2|$ with $\lambda/(2\pi)$. We limit ourselves to a subnetwork of wave packet revivals defined by two rational approximations of $1/\sqrt{2}$: $3/4$ and $5/7$. As we can see in the figure, the quality of self-reconstruction gradually fades with distance from the source because the error grows with $\mathrm{gcd}(p,q)$ for any given $q$. Compared to  $3/4$, $5/7$ yields a significantly smaller error in approximating $1/\sqrt{2}$, allowing high-quality revivals to persist over long propagation distances. In particular, our experiments demonstrate that the $5/7$-approximation is adequate to yield reliable replicas of the source up to the third order, which is entirely consistent with our theoretical predictions in Fig.\ref{fig2} (see also Fig.S6). Overall, the experimental observations in Figs.~\ref{fig3} and \ref{fig4} convincingly verify the existence of regular and irregular revival networks of quasi-periodic random wave packets.

\textit{Conclusions---}We have discovered theoretically and verified experimentally the existence of regular and irregular revival networks of the two-point autocorrelation function of an ensemble of quasi-periodic random waves with structured second-order correlations. We have also demonstrated that the ensemble average Talbot carpets of such random wave packets exhibit features that have no parallels among coherent wave packet revivals even for commensurable periods of the intensity and DOC at the source. 

We conjecture that the fidelity of our experimentally generated ensemble averaged replicas of the source can be further improved as follows. By comparing Eqs.~(\ref{Len}), ~(\ref{Len-in}) and~(\ref{Lss}), we notice that the ensemble self-imaging distances in either commensurable or incommensurable period cases form a subset of the distances over which the statistical stationarity (homogeneity) of the ensemble is recovered. Hence, if the ensemble is ergodic, the average over a number of ensemble realizations, which, in practice, introduces attenuation errors [c.f., Fig.~\ref{fig3}(b)\&(d)], can be replaced with a temporal (spatial) average over a single realization. The ergodicity of the ensemble can be ensured by wrapping an OCL into a slowly varying envelope such that the two-time (two-point) autocorrelation function of the source decays with time (spatial position) difference over the scales far exceeding the OCL period~\cite{papoulis1965random}. Such enveloped OCLs were realized experimentally in the spatial domain~\cite{chen2016experimental}. 

Our results are universally applicable to wave packets governed by the free-space Schr\"{o}dinger or paraxial wave equations, spanning atomic, matter, and optical waves. We also anticipate that similar revival phenomena can take place in self-imaging of spatial OCLs in graded-index media~\cite{ponomarenko2015self}, or matter wave coherence lattices in parabolic potentials.

\textit{Acknowledgments---}The authors acknowledge financial support from National Key Research and Development Program of China (2022YFA1404800); National Natural Science Foundation of China (12534014, 12374311, 12192254, W2441005); Taishan Scholars Program of Shandong Province (tsqn202312163); Natural Science Foundation of Shandong Province (ZR2025ZD21,ZR2023YQ006); Key research and development program of Shandong Province (2024JMRH0105) and the Natural Sciences and Engineering Research Council of Canada (RGPIN-2025-04064). 

\textit{Data availability---}The data that support the findings of
this article are not publicly available. The data are available
from the authors upon reasonable request.


\section{End Matter}
\renewcommand{\theequation}{A\arabic{equation}}
\setcounter{equation}{0}
Consider a two-time autocorrelation function  $\Gamma$ of the ensemble $\{\Psi\}$, defined as 
    \begin{equation}\label{G}
        \Gamma(\tau_1,\tau_2,z)=\langle \Psi^* (\tau_1,z)\Psi(\tau_2,z)\rangle.
    \end{equation}
On substituting from Eq.~(\ref{APS}) into~(\ref{G}) and averaging over the ensemble with the aid of Eq.~(\ref{stat}), we obtain, after elementary algebra, for the autocorrelation function the expression
\begin{equation}\label{Gam}
    \begin{split}
    \!\Gamma(&\tau_1,\tau_2,z)\!=\!\sum_{m,n,l}\!c_l^*c_n\lambda_m e^{-2\pi i(n\tau_2-l\tau_1)/T_p}e^{-2\pi im(\tau_2-\tau_1)/T_c} \\
    &\times\exp\left[-2\pi^2 i\beta_2z\frac{(n^2-l^2)}{T_p^2}\right]\exp\left[- 4\pi^2 i\beta_2z\frac{m(n-l)}{T_pT_c}\right].
    \end{split}
\end{equation}
Here $(n,m,l) \in\mathbb{Z}$. It follows from Eq.~(\ref{Gam}) that although the fields at the source are statistically stationary, they are, in general, nonstationary at $z=const\geq 0$. The statistical stationarity is recovered whenever the last phasor on the right-hand side of Eq.~(\ref{Gam}) attains unity, which occurs at 
\begin{gather}\label{Lss}
  z_s=\frac{sT_cT_p}{2\pi|\beta_2|}; \quad s\in\mathbb{N}.
\end{gather}

\end{document}